\documentclass[a4paper,11pt]{article}
\pdfoutput=1 
\usepackage{jcappub} 
\usepackage[T1]{fontenc}

\title{\boldmath{A possible solution to the Hubble tension from quantum gravity}}

\author{Anupama B}
\author{and P K Suresh}

\affiliation{School of Physics \\University of Hyderabad,\\P.O. Central University, Hyderabad - 500046, India.}

\emailAdd{21phph19@uohyd.ac.in}
\emailAdd{sureshpk@uohyd.ac.in}

\abstract{We investigate the relevance of quantum gravity during inflation to address the Hubble tension that arises from Planck 2018 and SH0ES data sets. We show that the effect of quantum gravity during inflation can increase the rate of change of $H_0$, thereby accounting for a wide range of observed $H_0$. Further, we show that due to the quantum gravity effect on inflation, the temperature at the onset of reheating can be lower than the standard case, causing delays in the reheating process. The role of quantum gravity is inevitable in settling the Hubble tension. The results of the present study may find use in resolving the Hubble tension, in validating inflationary model and quantum gravity.}

\begin{document}
\maketitle
\flushbottom

\section{Introduction}

The twentieth century discovery of expansion of the universe by Edwin Hubble has tremendously improved the understanding of the universe in a great way and has helped in strengthening the foundations of modern observational cosmology. This historical discovery is not complete without considering the contributions of Henrietta Leavitt, Vesto Slipher and George Lema\^{i}tre whose works were detrimental in developing the distance ladder method for measuring intergalactic distances \cite{henri}, finding the recessional velocities of distant galaxies \cite{vesto} and associating it with expansion of the universe \cite{lemaitre}. All these efforts along with the observations made by Hubble led to the Hubble-Lema\^{i}tre law \cite{iau}, a linear relation between the radial velocity of galaxies and their distances using the present value of the Hubble parameter ($H_0$) \cite{hubble}.  Hubble's initial estimate for $H_0$ was very high ($\approx$ 500 Km s$^{-1}$ MPc$^{-1}$)  due to errors in the calibration.\\

$H_0$ can be estimated by employing many modern techniques, which include the method of CMB and calibration of distance ladder using standard rulers \cite{devalentino}. The value of $H_0$ obtained from CMB calibrated observations of Planck 2018 in the light of $\Lambda$CDM model is 67.36 $\pm$ 0.54  Km s$^{-1}$ MPc$^{-1}$ \cite{aghanim} and the SH0ES (Supernova $H_0$ for the Equation of State) team has estimated the late universe value of $H_0$ as 74.03  $ \pm$  1.04  Km\  s$^{-1}$ \  MPc$^{-1}$ \cite{shoes}, which disagree to a level of 5$\sigma$. This incompatibility between the indirect measurements and the direct measurements of $H_0$ is known as the Hubble tension. In spite of the improvements in the detection mechanisms, the discrepancy still exists, pointing towards a growing tension. At present the disagreement is statistically significant creating a real tension among the cosmologists. Resolving the Hubble tension has paramount importance in cosmology as the value of $H_0$ plays a crucial role in bridging the theory and observations related to the determination of size, age and expansion rate of the universe.\\

With the developments in precision cosmology and the modern methods of astronomical observation and techniques, we can assume that both the large scale and local measurements of $H_0$ are reliable. In view of this, many attempts have been made to resolve the Hubble tension such as taking into account the effect of local inhomogeneity \cite{ldl}, early dark energy models \cite{ede}, modified gravity theory \cite{jordan,serg}, phantom cosmology (w $< $ $-$ 1) and the increase in the number of relativistic degrees of freedom due to effective operators \cite{deg} and vacuum energy interaction with matter and radiation \cite{vacint}. New measurements of $H_0$ from the shadows of supermassive black holes \cite{ssmb}, grey sirens \cite{gs} and ellipsoidal geometry of the universe are also in progress \cite{ellipse}. While one set of researchers argues that the problem lies in the assumptions made in the standard $\Lambda$CDM model of cosmology others are searching for new Physics. \\

Often, dark energy is cited as responsible for new Physics. But, a recent study using Pantheon+ data to understand the cosmological foundations is now gaining attention \cite{35} which compels us to focus on the mathematical side of general relativity. Further, the cosmological principle is itself an assumption and therefore the ideal Friedmann-Lema\^{i}tre-Robertson-Walker (FLRW) metric may not be complete to address the tensions in cosmology \cite{nf}. Since $\Lambda$CDM model is based on cosmological principle and FLRW metric, there are suggestions for a model-independent approach to study cosmology with non-FLRW evolution \cite{ct}. Therefore the new Physics could also be influenced by the interaction between matter and non-linear geometry of spacetime \cite{35}. This demands the necessity of timescape cosmology that deals with the origin of scale and synchronisation of various cosmic clocks \cite{ts}. However, the present study does not address such issues.\\

We assume that at very early stage, the universe underwent an accelerated expansion in a very short span of time known as inflation. Usually, the inflation is described by a solo scalar field. The variations in this field can have energy higher than the Planck scale ($m_{pl}$), the energy scale where the quantum gravity effect is supposed to be dominant as implied by the Lyth bound \cite{lyth}
\begin{equation}
\Delta \varphi \gtrsim m_{pl}\sqrt{\frac{r}{4\pi}} .
\end{equation}Here, $r$ represents the ratio of the tensor power spectrum to that of the scalar power spectrum arising from the primordial perturbations associated with the inflation calculated at a pivot scale.\\

The recent CMB anisotropic measurements rule out most of the single scalar field based inflation and multi fields inflationary model can be adopted as an alternate remedy to it. Among the multi fields inflationary models, hybrid inflation receives much attention in explaining the gap between inflation and particle production \cite{linde}. The recent study on hybrid inflation shows that the Hubble parameter during different epochs of evolution of the universe can be related \cite{bourakadi}. Therefore in the present work, we attempt to address the Hubble tension using a hybrid inflationary model by incorporating quantum gravity effect in the framework of effective field theory (EFT). We investigate the Hubble parameter during the inflation ($H_I$) and during the phase transition ($H_T$) in terms of the quantum gravity effect in view of Planck 2018 and SH0ES data. We show that quantum gravity influences $H_I$ whereas it does not affect $H_T$. Therefore we explore the possibility of accounting a wide range of observed $H_0$ values from various estimates. Since quantum gravity is sensitive to inflation in the EFT framework its reflection is expected on the immediate stage like reheating. This impact on $H_I$ can be tested by analysing its consequence on reheating by examining the reheating e-folding number and reheating temperature. We show that the temperature during the onset of reheating can be lowered due to quantum gravity resulting in delayed reheating. Throughout the paper we follow c = G = $\hbar$ =1.
\\


\section{The Hybrid Inflationary Model}

The concept of inflation was introduced to resolve the major problems of standard cosmology like flatness problem, horizon problem, fine tuning problem, etc \cite{guth}. More than hundred models have been proposed so far but single field models alone are not sufficient to address all the problems of the standard model of cosmology successfully along with various observational results \cite{encyclo}. Multi fields models are introduced to overcome this issue. Among these models, the hybrid inflationary model consisting of two scalar fields ($\varphi$, $\sigma$) is considered reasonable to study the inflation, where the slow roll of the field ($\varphi$) is responsible for inflation and the waterfall field ($\sigma$) leads to symmetry breaking and phase transition \cite{linde}.\\
\begin{figure}[t]
   \centering
   \includegraphics[width=0.8\linewidth]{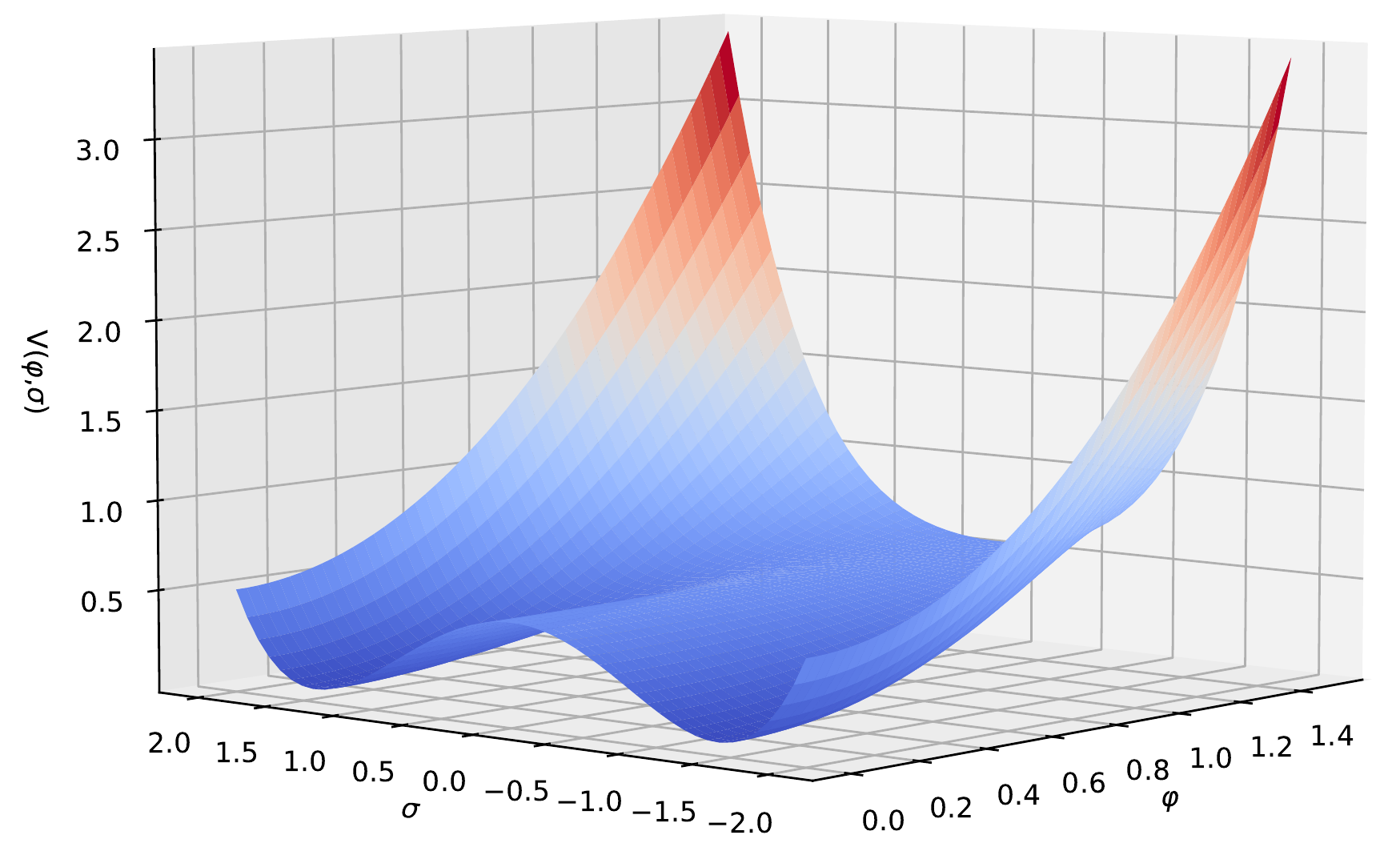}
      \caption{Schematic representation of hybrid inflationary potential.}
   \label{pot}
\end{figure}

The potential of hybrid inflationary model is given by \cite{linde} (see figure (\ref{pot}))
\begin{equation}\label{bj91}
V(\varphi, \sigma ) = \frac{1}{4\lambda} \bigg(M^2 - \lambda \sigma^2 \bigg)^2 + \frac{1}{2}b^2 \varphi^2 \sigma^2 + \frac{1}{2}m^2 \varphi^2,
\end{equation}
where $M$ and $m$ are the masses of  $\sigma$ and $\varphi$ respectively, $\lambda$ is the self coupling constant for $\sigma$ and $b$ is the coupling constant for the interaction between $\varphi$ and $\sigma$.
The critical value of the field $\varphi$ is found to be $\varphi_c$ $=$ $\frac{M}{b}$. In the hybrid model, the mechanism of inflation begins when $\varphi$ $>$ $\varphi_c$ which is followed by the slow roll of $\varphi$ towards $\varphi_c$. At $\varphi$ $ =$ $ \varphi_c$, the inflation comes to an end. When $\varphi$ $<$ $\varphi_c$ (waterfall regime) \cite{waterfall}, symmetry breaking occurs and both the fields roll down to their respective true vacuum  $\varphi$ $\rightarrow$ 0 and $\sigma$$ \rightarrow$ $\frac{M}{\sqrt{\lambda}}$ rapidly.\\

\begin{figure}[t]
   \centering
   \includegraphics[width=0.6\linewidth]{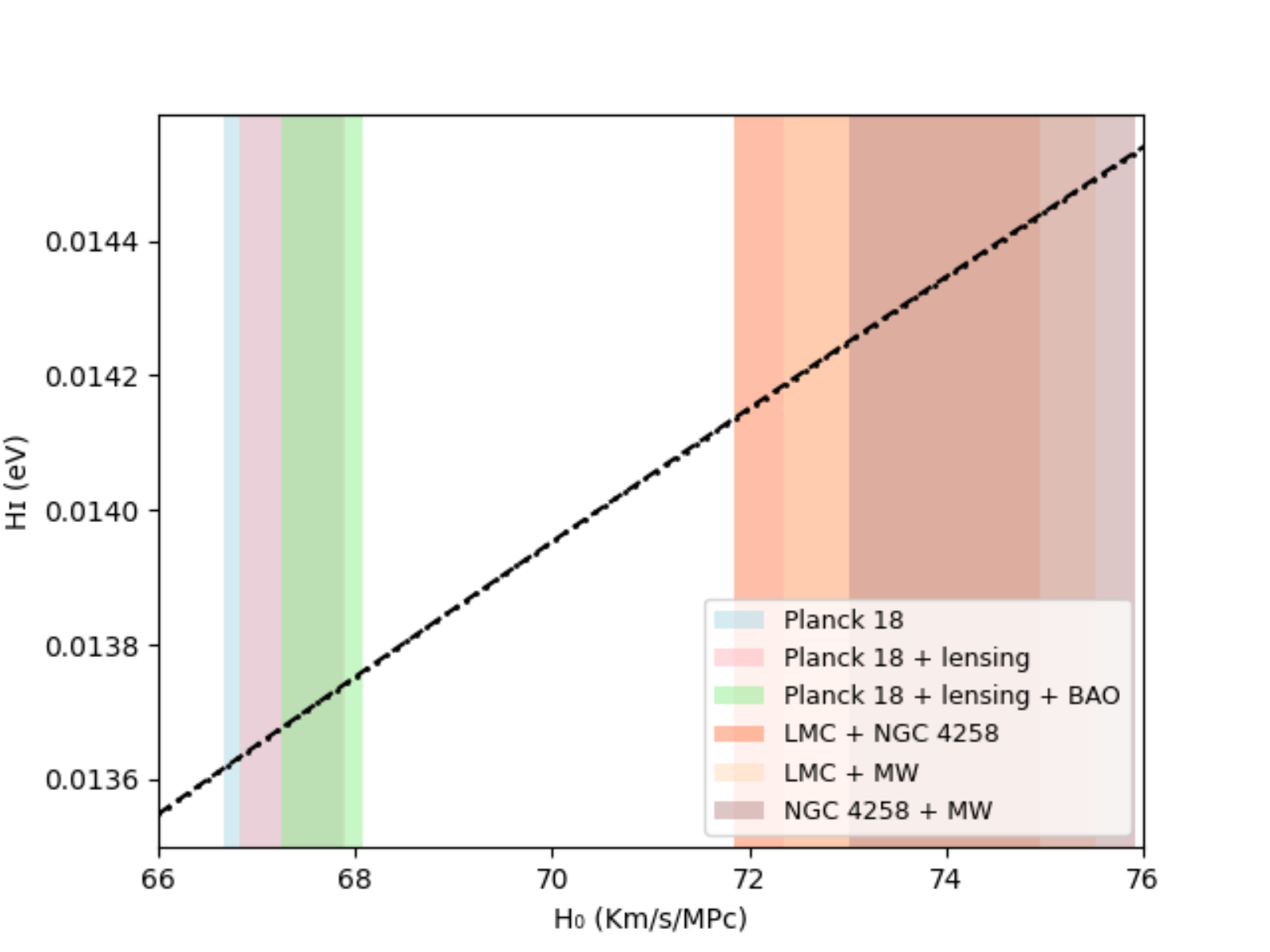}
     \caption{Variation of the Hubble parameter during inflation ($H_I$) for a range of present Hubble parameter ($H_0$) with Planck 2018 and SH0ES data.}
   \label{HI}
\end{figure}

During the initial stage of inflation ($\varphi$ $>$ $\varphi_c$ )  the potential is nearly flat in the direction of the field $\varphi$ and is steeper in the direction of $\sigma$. As a result, the field $\sigma$ immediately settles in the false vacuum ($\sigma$ $=$ 0) and $\varphi$ rolls slowly subjected to quantum fluctuations. 
Therefore $\varphi$ can be written as 
\begin{equation}
\varphi(x,t) = \varphi(t) + \delta \varphi(x,t) .
\end{equation}
The magnitude of these fluctuations in the Fourier space is given by \cite{bdv}
\begin{equation}
| \delta \varphi_k |^2 \ \simeq \ \frac{H_{\tiny I}^2}{2k^3} \bigg( \frac{k}{aH_{\tiny I}} \bigg)^{ \frac{2m^2}{3H_I^2}} ,
\end{equation}
and the variance of the scalar field fluctuations is given by 
\begin{equation}\label{bj90}
<|\delta \varphi_k|^2 > \  \ \simeq \  \frac{3H_I^4}{8\pi^2m^2} \bigg[ 1 - e^{- \frac{2m^2N}{3H_I^2}} \bigg] ,
\end{equation}
where $H_I$ is the Hubble parameter during inflation, $k$ is the wavenumber for the mode that exits the horizon and $N$ is the e-folding number that amounts the duration of inflation. Inflation occurs for a long period ($N$ is very high) therefore equation (\ref{bj90}) becomes
\begin{equation}\label{fluct}
<|\delta \varphi_k|^2 >  \ \ \simeq \   \frac{3H_I^4}{8\pi^2m^2} .
\end{equation}
Since the inflation is governed by the field $\varphi$, the relevant potential can be written as
\begin{equation}\label{orinf}
V(\varphi) = \frac{M^4}{4 \lambda} + \frac{1}{2}m^2 \varphi^2 .
\end{equation}
The quantum fluctuations can dominate over homogenous part of the field. Therefore using equation (\ref{fluct}) in (\ref{orinf}) the potential can be rewritten as
\begin{equation}\label{bj}
V(\varphi) \  \simeq \ \frac{1}{2}m^2  <|\delta \varphi_k|^2 > \ \simeq   \ \frac{3H_I^4}{16\pi^2} .
\end{equation}
Here, we do not distinguish the behaviour of early dark energy and late dark energy \cite{ups,kami} but assume that the early dark energy component evolves to become the present cosmological constant ($\Lambda$) with fractional energy density $ \Omega_{\Lambda}$.  With this understanding, we compute the Hubble parameter during inflation using the equation (\ref{bj}) in Friedmann's equation for a flat FLRW universe and obtain \cite{bourakadi}
\begin{equation}\label{bj21}
H_I = (\Omega_{\Lambda})^{\frac{1}{4} } \sqrt{4 \pi m_{pl} \ H_0 } .
\end{equation}
We study the obtained Hubble parameter during inflation for a range of values of the present Hubble parameter estimated from Planck 2018 and SH0ES data. The results are presented in figure (\ref{HI}). 

\subsection{Phase transition}
\begin{figure}[t]
   \centering
   \includegraphics[width=0.6\linewidth]{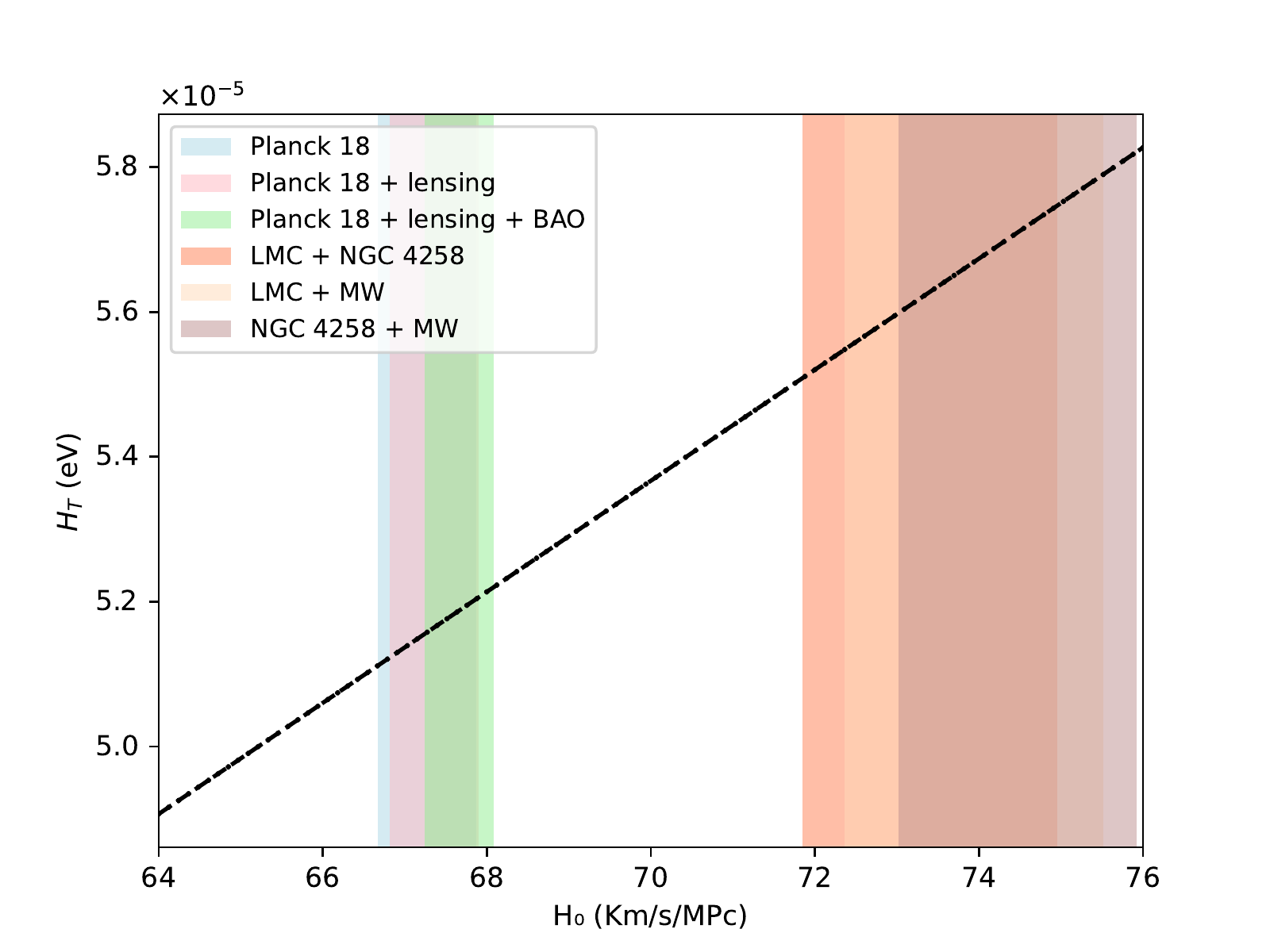}
   \caption{{Variation of the Hubble parameter during phase transition ($H_T$) for a range of present Hubble parameter ($H_0$) with Planck 2018 and SH0ES data.}}
   \label{HT}
\end{figure}

The initial conditions of inflation are chaotic but as soon as $\varphi$ $<$ $\varphi_c$,  ordering becomes more and symmetry becomes less. In hybrid inflation, this happens instantaneously through waterfall mechanism known as phase transition and the symmetry is said to be broken. The system settles down in the lowest energy state (true vacuum) of the fields. The hybrid inflation results in spontaneous symmetry breaking and first order phase transition, driven by the vacuum energy density \cite{linde}. Therefore the potential during the phase transition is
\begin{equation}\label{bj2}
V_T = V(0,0)= \frac{M^4}{4 \lambda} \simeq \frac{H_T^2}{4} ,
\end{equation}
where $H_T$ is the Hubble parameter during phase transition. Again, we assume that there is no distinction between early dark energy and late dark energy. Therefore, $H_T$ can be obtained using equation (\ref{bj2}) in Friedmann's equation for a flat FLRW universe as
\begin{equation}\label{bj20}
H_T = \sqrt{12  m_{pl}^2 \Omega_{\Lambda}} \ H_0 .
\end{equation}
We study the Hubble parameter during phase transition for a range of values of the present Hubble parameter obtained from Planck 2018 and SH0ES data. The results are presented in figure (\ref{HT}).

\section{Reheating}
\begin{figure}[t]
   \centering
   \includegraphics[width=0.6\linewidth]{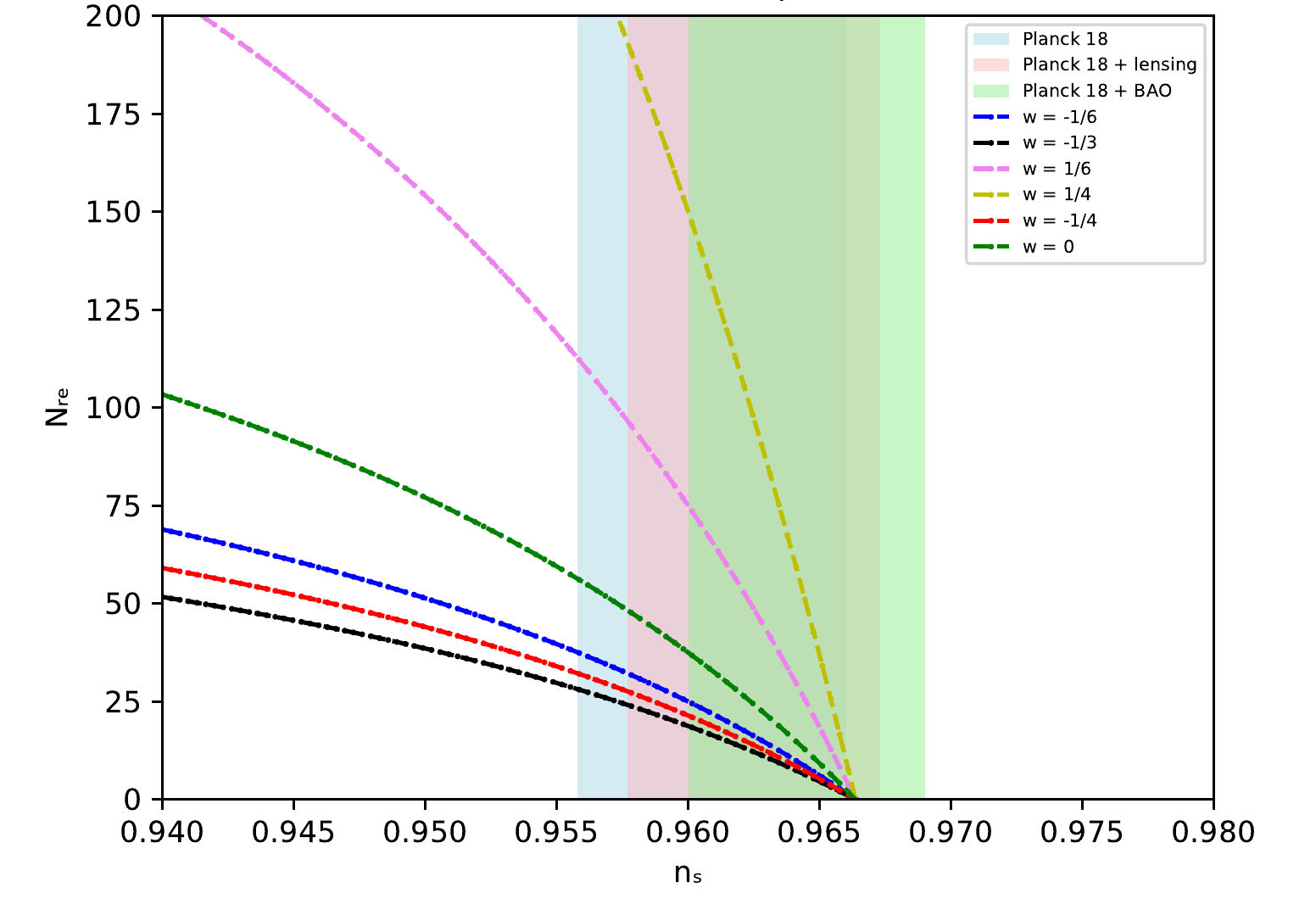}
      \caption{Behaviour of the reheating e-folding number ($N_{re}$) for a range of scalar spectral index ($n_s$) for various equation of state (w) with Planck 2018 data.}
   \label{Nre}
\end{figure}

\begin{figure}[t]
   \centering
   \includegraphics[width=0.6\linewidth]{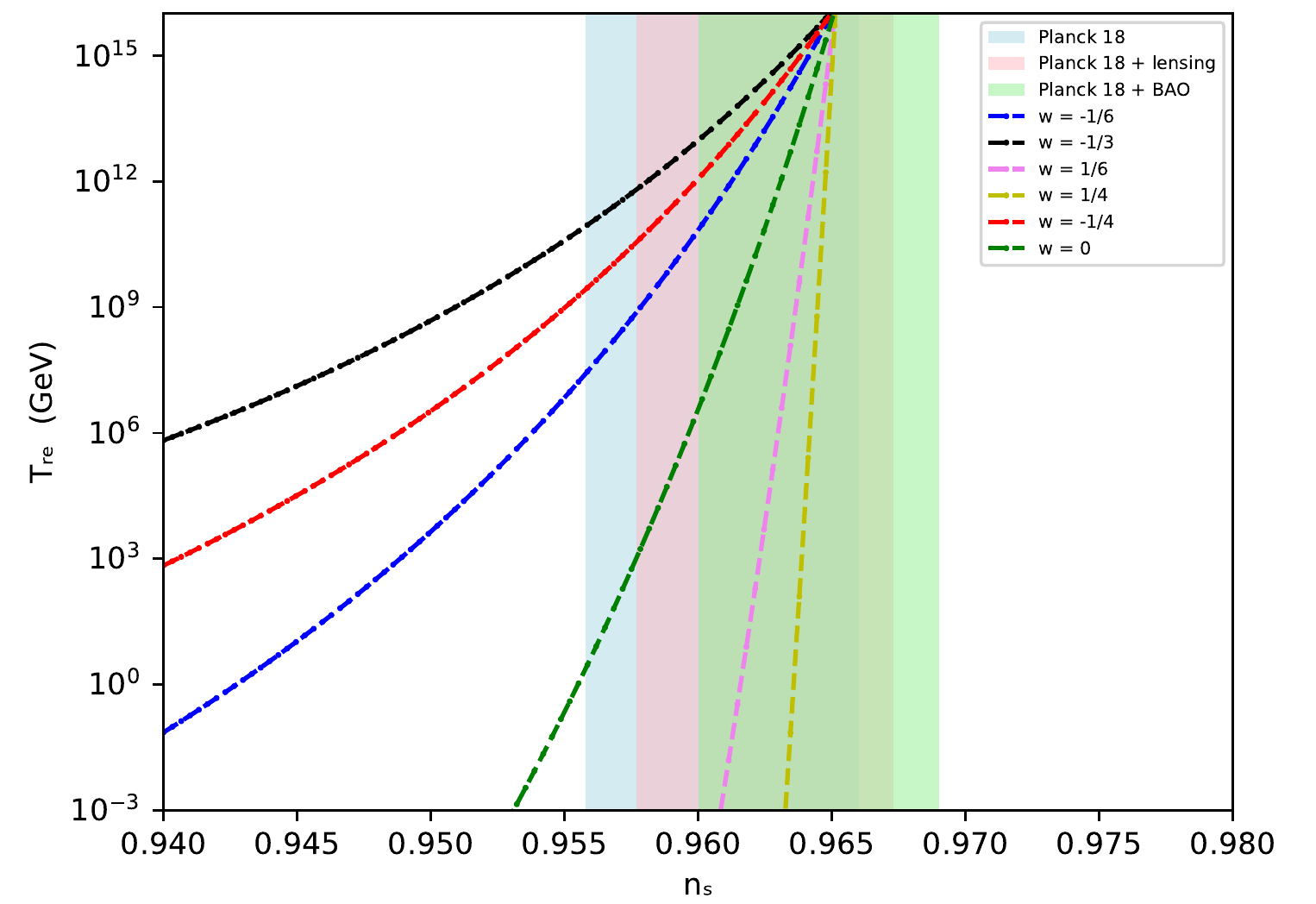}
      \caption{Behaviour of the reheating temperature ($T_{re}$) for a range of scalar spectral index ($n_s$) for various equation of state (w) with Planck 2018 data.}
   \label{Tre}
\end{figure}

At the end of inflation, the universe was devoid of any matter and further the temperature dropped down to a level that was not sufficient to trigger the big bang nucleosynthesis. To initiate the thermonuclear reactions, temperature of the universe must be greater than 10 MeV \cite{akhilesh,kaz1,kaz2}. In order to achieve this threshold temperature a mechanism known as reheating is required. Reheating is the phase between the end of inflation and the beginning of radiation dominated epoch where the energy density of the inflaton field is transferred to the matter and thermalization occurs. The dynamics of the energy density of a cosmological fluid is governed by the time dependent effective equation of state (w). It can be used to define the two important parameters of reheating, the reheating e-folding number ($N_{re}$) and the reheating temperature ($T_{re}$). The initial stage of reheating where the scalar particles are produced is called preheating and can be described using two scenarios. One is the perturbative decay of the inflaton as it oscillates near the minimum of its potential \cite{historical} and the other scenario involves a large amount of particle production via non perturbative processes like tachyonic instability \cite{tachyonic}, parametric resonance \cite{parametric,parametric2} and instant preheating \cite{instant}. Effective preheating is necessary to achieve the required reheating temperature. The duration of preheating is expressed in terms of the preheating e-folding number $(N_{pre})$  \cite{gauss} and the e-folding number for thermalization ($N_{th}$). Preheating and thermalization together constitute reheating and can be written as
\begin{equation}
N_{re} = N_ {pre} + N_{th} ,
\end{equation}
where 
\begin{equation}
N_{pre} = \bigg[ 61.6 - \frac{1}{4} \ln \bigg( \frac{V_{end}}{H_k^4} \bigg) - N \bigg] - \bigg( \frac{1-3\text{w}}{4} \bigg)N_{th} .
\end{equation}
In the present work we assume instantaneous preheating ($N_{pre} \rightarrow 0$ ), which gives the reheating e-folding number as
\begin{equation}\label{bj3}
N_{re} = N_{th} = \frac{4}{1-3 \text{w}}\bigg[ 61.6 - \frac{1}{4} \ln \bigg( \frac{V_{end}}{H_k^4} \bigg) - N \bigg],
\end{equation}
where $V_{end}, H_k$ and $N$ can be written in terms of the scalar spectral index ($n_s$) and amplitude of the scalar power spectrum ($A_s$)  \cite{akhilesh,kosar,n_s} as
\begin{eqnarray}
\label{bj7}
V_{end} &=& 6 \pi^2 m_{pl}^2 A_s \bigg( \frac{1-n_s}{2} \bigg)^2 \\ [0.5cm]
\label{bj8}
H_{k} &=& 2 \pi \sqrt{A_s \ m_{pl}^4 \bigg( \frac{V'}{V} \bigg)^2} \\ [0.5cm]
\label{bj9}
H_{k} &=& \frac{\pi^2 \ m \ m_{pl}^2}{H_I^2} \sqrt{\frac{128 \ A_s}{3}} \\ [0.5cm]
\label{bj10}
N &=& \frac{2}{1-n_s}.
\end{eqnarray}
By substituting equations (\ref{bj7}), (\ref{bj9}) and (\ref{bj10}) in equation (\ref{bj3}) we get
\begin{equation}\label{N_re}
N_{re} = \frac{4}{1-3 \text{w}}\bigg[ 61.6 - \frac{1}{4} \ln \bigg( \frac{6912 \  \Omega_{\Lambda}^2  \ (1-n_s)^2}{32768 \ \pi^2 \ A_s} \bigg) - \frac{2}{1-n_s} \bigg].
\end{equation}
Soon after the particle production at the end of inflation, thermalization occurs through various mechanisms like back reaction and rescattering. The universe reaches a thermal equilibrium by attaining a temperature ($T_{re}$) and the reheating comes to an end. This reheating temperature can be expressed in terms of the degree of freedom of the relativistic species ($g_*$) as \cite{akhilesh,kosar,n_s}
\begin{equation}\label{bj11}
T_{re} = \bigg[ \bigg(\frac{43}{11g_*} \bigg)^\frac{1}{3} \frac{a_0 \ T_0 \ H_k \ e^{-N}}{k} \bigg(  \frac{45 \ V_{end}}{\pi^2g_*} \bigg)^\frac{-1}{3(1+\text{w})} \bigg]^\frac{3(1+\text{w})}{3\text{w}-1}.
\end{equation}
In order to study $T_{re}$ with Planck 2018 data it is convenient to express $T_{re}$ in terms of $n_s$ and $A_s$. Therefore we rewrite equation (\ref{bj11}) using equations (\ref{bj7}),  (\ref{bj9}) and (\ref{bj10}) as follows 
\begin{eqnarray}\label{T_re}
T_{re} &=& \bigg(\frac{43}{11}\bigg)^{\frac{1+\text{w}}{3\text{w}-1}} \ \bigg( \frac{135}{2} \bigg)^{\frac{-1}{3\text{w}-1}}  \ {g_*}^{\frac{-\text{w}}{3\text{w}-1}} \ \bigg( \frac{\sqrt{2} \ \pi \  a_0 \ T_0 }{k} \bigg)^\frac{3(1+\text{w})}{3\text{w}-1} \ m_{pl}^{\frac{3\text{w}+1}{3\text{w}-1}} \\ \nonumber
&& \left. \right. \\ \nonumber
&& \left.  \times \  A_s^{\frac{1+3\text{w}}{2(3\text{w}-1)}} \ (1-n_s)^{\frac{1}{2}} \  e^{\frac{-6(1+\text{w})}{(1-n_s)(3\text{w}-1)}}  \right. .
\end{eqnarray}
Now we are in a position to study the reheating e-folding number and reheating temperature for a range of scalar spectral index obtained from Planck 2018 data for different equation of state. The results are presented in figure (\ref{Nre}) and (\ref{Tre}) respectively.


\section{Effective Field Theory and Quantum Gravity }

The Lyth bound \cite{lyth} suggests that the inflationary energy scale can be much more than the Planckian energy scale where quantum gravity is relevant. Therefore quantum gravity effect cannot be ignored while explaining the Physics of early universe and hence the inflationary scenario. But this effect is accessible only at high energy scales and its direct realization is difficult. Even though many attempts have been made to formulate quantum gravity, so far no satisfactory theory exists. The effective field theory (EFT) approach is a promising one to probe the quantum gravity effect through inflation \cite{donos,calmet}.\\ 

In the context of EFT \cite{excursion}, the Einstein-Hilbert action for the hybrid inflationary model can be written as
\begin{eqnarray}
S &=& \int d^Dx \ \sqrt{-g} \bigg( \frac{m_{pl}^2R}{2} +f(\varphi,\sigma) F(R,R_{\mu \nu})+ g^{\mu\nu}\partial_\mu \varphi \partial_\nu \varphi \\ \nonumber
&& \left. \right. \\ \nonumber
&& \left. + \frac{1}{4\lambda} \bigg(M^2 - \lambda \sigma^2 \bigg)^2   + \ \frac{1}{2}b^2 \varphi^2 \sigma^2 + \frac{1}{2}m^2 \varphi^2 + \sum_{n=5}^{\infty} c_n \frac{\varphi^n}{m_{pl}^{n-4}} \right. \bigg) ,
\end{eqnarray}
where $f(\varphi,\sigma) F(R,R_{\mu \nu})$ represents the non minimal coupling of the fields to gravity and $c_n \frac{\varphi^n}{m_{pl}^{n-4}}$ are called as the higher dimensional operators (HDO) where $c_n$ are the Wilson coefficients of HDO whose values must be of the order of $10^{-3}$ for the potential to be nearly flat and the slow roll to occur. In the current work, we study the hybrid inflation, where, the field $\varphi$ governs the inflation and the second field $\sigma$ is subdominant during inflation therefore it plays a role only during phase transition. Hence for simplicity, in the present study we assume $\varphi$ is minimally coupled to gravity and the second term can be ignored (for details see \cite{excursion}). Also, for simplicity, we neglect the term containing partial derivatives. Since the inflation is governed only by $\varphi$, the effective potential of inflation becomes,
\begin{equation}
V_{eff}(\varphi) = V_{ren}(\varphi) + \sum_{n=5}^{\infty} c_n \frac{\varphi^n}{m_{pl}^{n-4}}.
\end{equation} 
Here, it is assumed that $V_{ren}(\varphi)$ contains renormalizable terms of the potential up to the relevant dimension.


\subsection{Quantum gravity effect on Hubble parameter}
In view of EFT, the effective potential corresponding to the inflationary field of the hybrid model with quantum gravity correction can be written as 

\begin{equation}
V_{eff}(\bar\varphi) = m_{pl}^4 (\bar m^2 \bar\varphi^2 + c_n \bar\varphi^n),
\end{equation}
where  $\bar m = \frac{m}{m_{pl}}$ and $\bar \varphi=\frac{\varphi}{m_{pl}}$ are the rescaled mass and field. For any model of inflation, all the associated Wilson coefficients need not be considered. But in the present study, we restrict ourselves to one dominant coefficient, say $c_6$  and hence the other coefficients are considered subdominant. Since the principal term of the potential is $\bar m^2$, the quantum gravity correction term can be taken as
\begin{equation}\label{bj13}
c_6 = \alpha_m \bar m^2 .
\end{equation}
Using equation (\ref{bj13}) the potential can be rewritten as
\begin{equation}\label{bj5}
V_{eff}(\bar\varphi) =  m_{pl}^4 \bar m^2 \bar\varphi^2(1+\alpha_m \bar\varphi^4),
\end{equation}
such that
\begin{equation}
|\alpha_m| \bar\varphi^4 < 1.
\end{equation}

Using the condition to end the inflation, the value of the rescaled inflationary field at the beginning of inflation in terms of the e-folding number with quantum gravity correction is \cite{excursion} 
\begin{equation}\label{bj4}
\begin{split}
\bar \varphi_N^2 &= \bar\varphi^2_{N , cl} + \bar\varphi^2_{N , qg}  \\ &= \frac{N}{2 \pi} + \frac{  \alpha_m N^3 }{12\pi^3} .
\end{split}
\end{equation}
Hereafter, the subscripts $cl$ and $qg$ respectively represents the standard classical part and quantum gravity corrected part of the associated quantities. With the help of the effective potential (in equation (\ref{bj5})) the slow roll parameters can be computed \cite{excursion}. Following equation (\ref{bj4}) the quantum gravity corrected first slow roll parameter ($\epsilon$) and second slow roll parameter ($\eta$) in terms of $N$ and $\alpha_m$ are \cite{excursion,sur}
\begin{equation}\label{eps}
\begin{split}
\epsilon &= \epsilon_{cl} + \epsilon_{qg} \\ &= \frac{1}{2N}  +  \frac{5 \alpha_m N}{12 \pi^2}, \\  \eta &= \eta_{cl} + \eta_{qg} \\ &=  \frac{1}{2N}  +  \frac{5 \alpha_m N}{3 \pi^2} \ .
\end{split}
\end{equation}
The scalar power spectrum arising from the scalar field fluctuations can be characterised in terms of the scalar spectral index and amplitude. Using equation (\ref{eps}) in the standard definition of  $n_s$ and $A_s$, the corresponding quantum gravity corrected results can be respectively written as
\begin{equation}
\label{nsqg}
\begin{split}
n_{s} &= n_{s, cl} + n_{s, qg} \\ &= \bigg(1 - \frac{2}{N} \bigg)+ \frac{5 \alpha_m N }{6 \pi^2}\  ,
\end{split}
\end{equation} 
\begin{equation}
\label{asqg}
\begin{split}
A_{s} &= A_{s, cl}+A_{s, qg} \\ &= \frac{4 m^2 N^2  }{3 \pi} + \frac{4 m^2 \alpha_m N^4 }{9\pi^3} \ .
\end{split}
\end{equation}
Next, we focus on the Hubble parameter during inflation in the light of quantum gravity. $H_I$ can be expressed in terms of slow roll parameter from the equations (\ref{bj8}) and (\ref{bj9}) as
\begin{equation}\label{bj6}
H_I = \bigg( \frac{16}{3 \epsilon} \bigg)^{\frac{1}{4}} \sqrt{\pi \ m \ m_{pl}} .
\end{equation}
Substituting equation (\ref{eps}) in equation (\ref{bj6}) the quantum gravity corrected $H_I$ is obtained as
\begin{equation}\label{bj14}
\begin{split}
H_{I} &= (\Omega_{\Lambda})^{\frac{1}{4}} \sqrt{4 \pi m_{pl} \ H_0 } \  \bigg( 1 - \frac{ 5 \alpha_m N^2 }{24 \pi^2} \bigg) .
\end{split}
\end{equation}
\begin{figure}[t]
   \centering
   \includegraphics[width=0.5\linewidth]{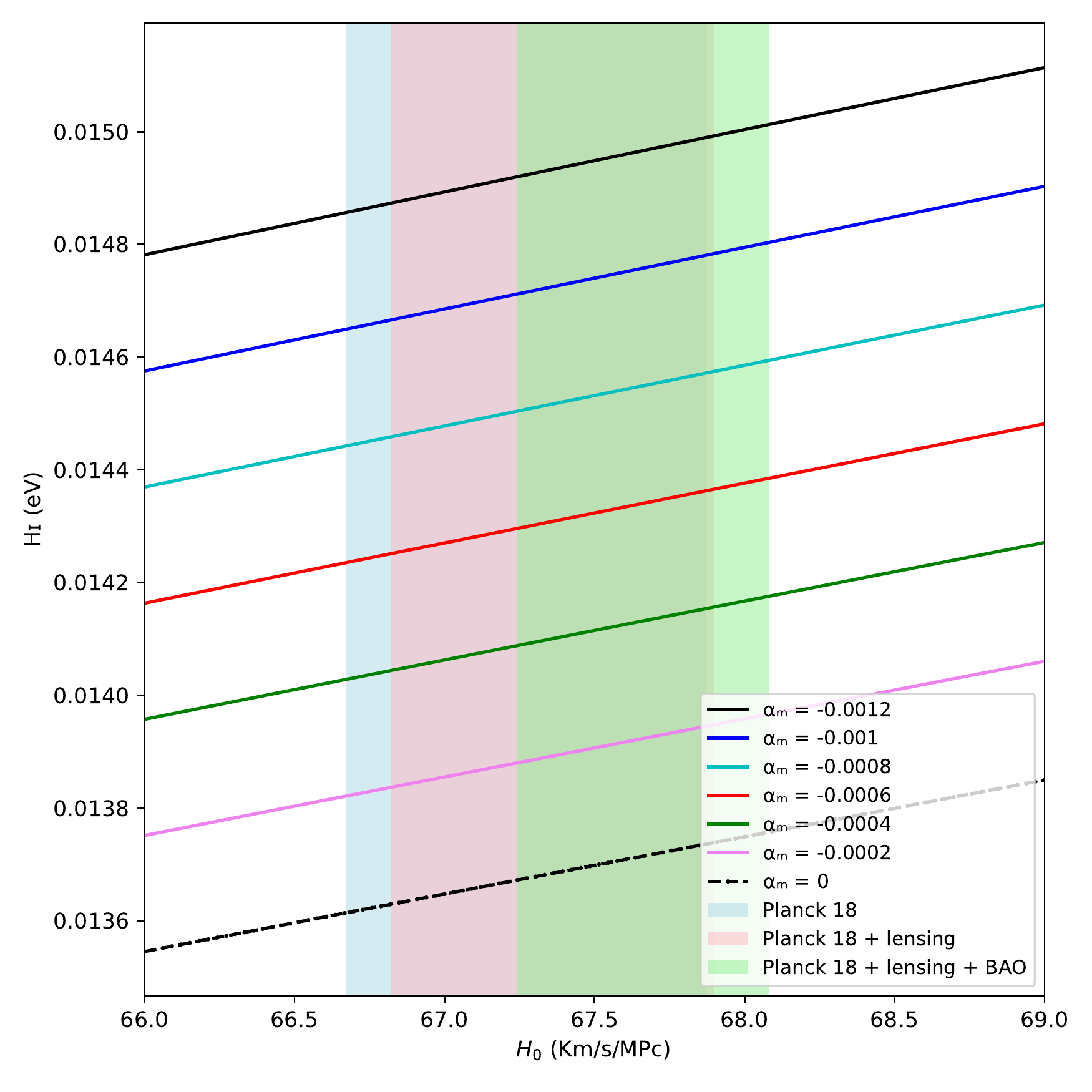}
   \caption{Effect of quantum gravity on the variation of the Hubble parameter during inflation ($H_I$) for a range of present Hubble parameter ($H_0$) with Planck 2018 data. }
   \label{diffalpha}
\end{figure}
\\
The quantum gravity corrected Hubble's parameter during inflation is studied for various measured values of $H_0$. The corresponding results are presented in figure (\ref{diffalpha}). At a glance, the figure shows that the quantum gravity corrected $H_I$  appears to be parallel to the classical counterpart. But equation (\ref{bj14}) demands a deeper analysis to understand the results further. Therefore to get more insight, we analyse the results in terms of the rate of change of $H_I$ with respect to the rate of change of $H_0$ for which we calculate the slope $(s)$ as 
\begin{equation}
s = \frac{d H_I }{d H_0} = ( \Omega_{\Lambda} )^{\frac{1}{4}}  \sqrt{ \frac{ \pi  m_{pl}}{H_0} } \ \bigg( 1- \frac{5 \alpha_m N^2}{24 \pi^2} \bigg).\label{S}
\end{equation}
\begin{figure}[t]
   \centering
   \includegraphics[width=0.5\linewidth]{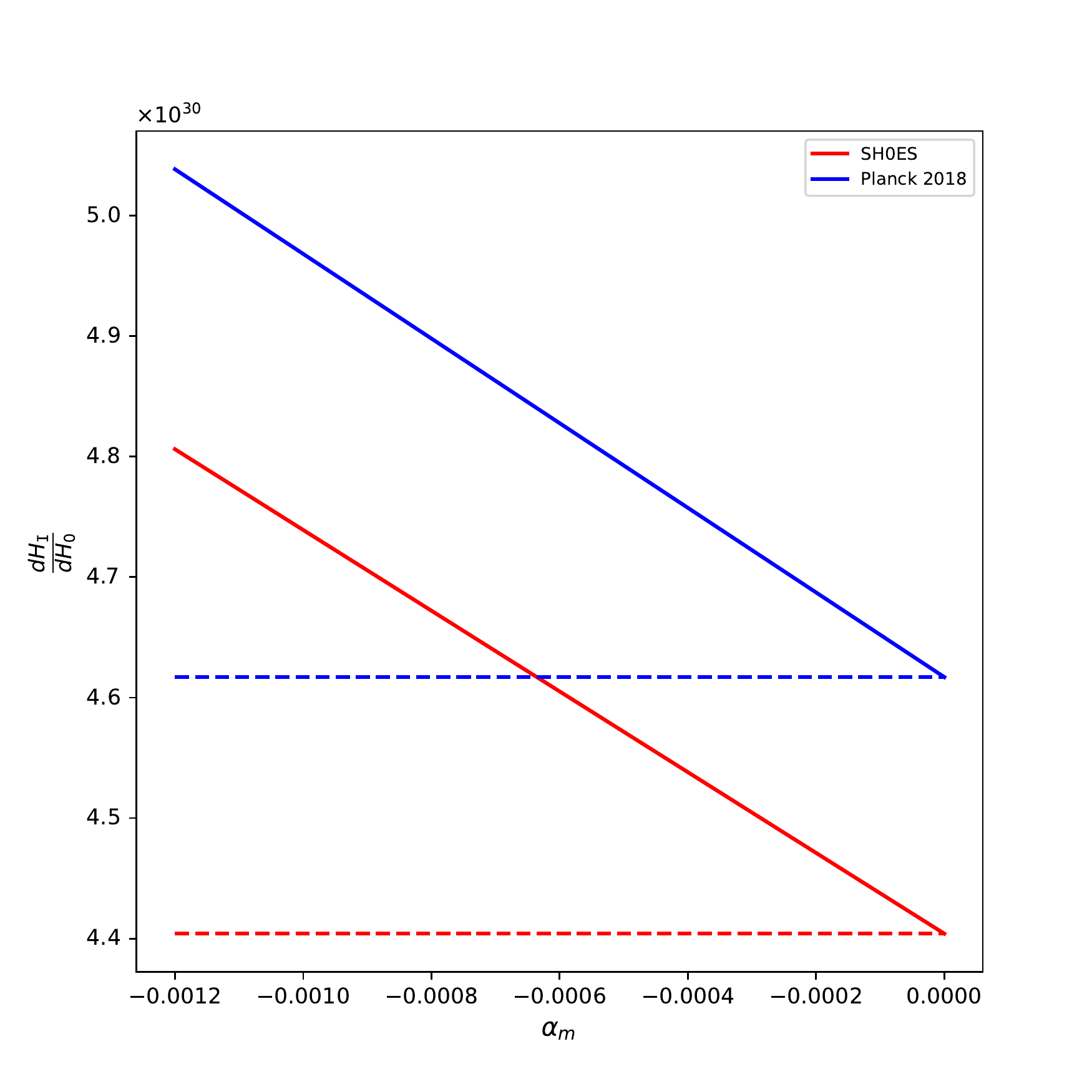}
   \caption{Quantum gravity effect on the slope ($s = \frac{dH_I}{dH_0}$) for a range of higher dimensional operator ($\alpha_m$) with Planck 2018 and SH0ES data. The blue and red dashed lines respectively represent the constant slope at $\alpha_m=0$, implying the absence of quantum gravity effect for Planck 2018 and SH0ES data.}
   \label{slopeHI}
\end{figure}
\\
We scrutinise the slope for various values of HDO using the Hubble parameter obtained from Planck 2018 and SH0ES data and the results are presented in figure (\ref{slopeHI}).
From the analysis of the results, the slope associated with quantum gravity corrected $H_I$ and $H_0$ is found decreasing.
Since $s$ is decreasing in nature we can infer that the rate of change of the Hubble parameter during inflation ($dH_I$) is decreasing due to the quantum gravity effect which in turn implies that the rate of change of the present Hubble parameter ($dH_0$) is increasing. In the absence of quantum gravity effect the HDO in equation (\ref{S}) vanishes and the slope remains unaffected (see figure (\ref{slopeHI})). Therefore we may conclude that the role of quantum gravity is important in understanding Planck 2018 measured $H_0$ while addressing the Hubble tension. 

\begin{figure}[t]
   \centering
   \includegraphics[width=0.5\linewidth]{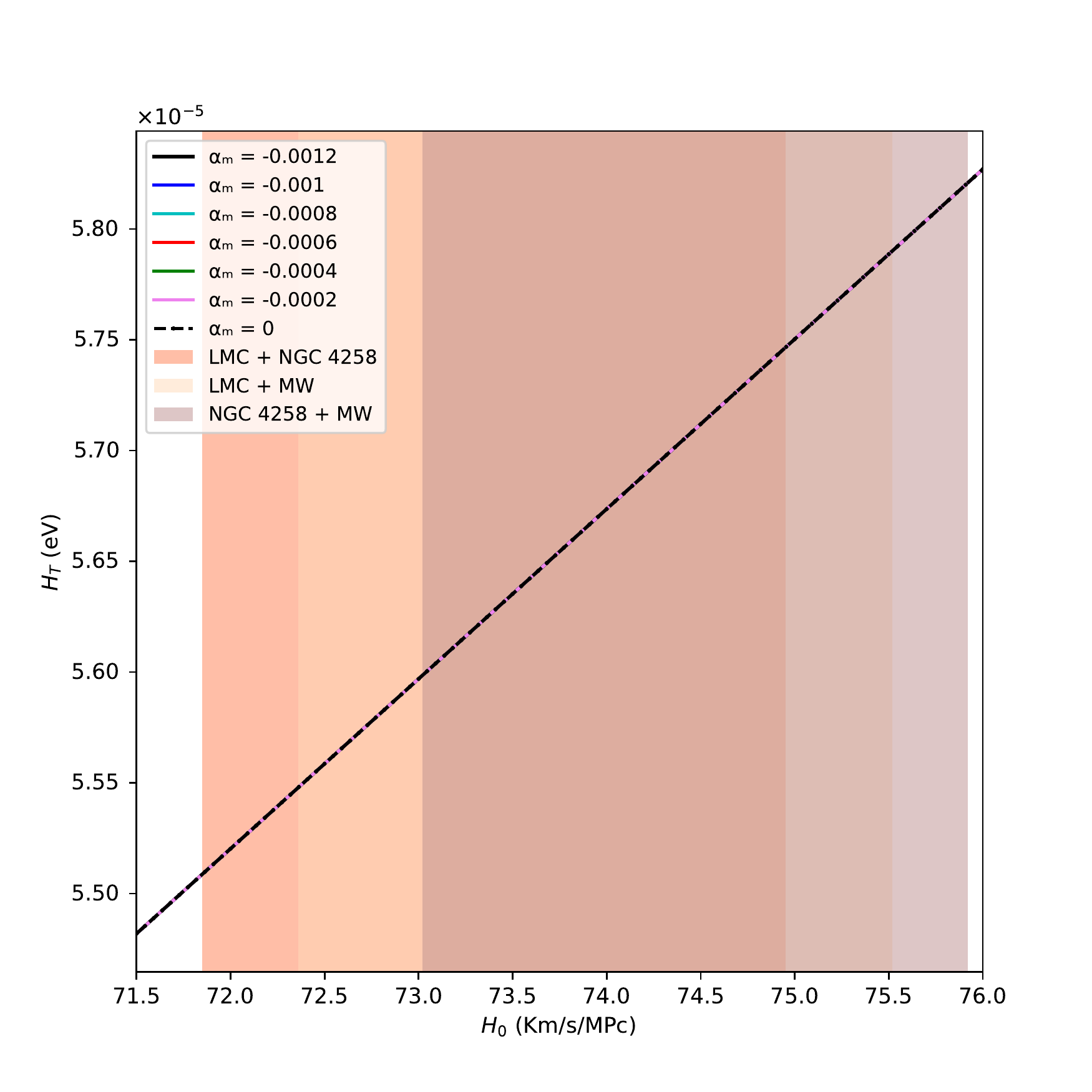}
      \caption{Variation of the Hubble parameter during phase transition ($H_T$) for a range of present Hubble parameter ($H_0$) with SH0ES data.}
   \label{HTdiffalpha}
\end{figure}

According to the hybrid inflationary model, soon after the inflation, the universe undergoes a phase transition driven by the vacuum energy density. The Hubble parameter during phase transition is given by equation (\ref{bj20}) and we can observe that $H_{T}$ is independent of HDO. This implies that
\begin{equation}\label{bj30}
\begin{split}
H_T &= H_{T, cl}  .
\end{split}
\end{equation}
We study $H_T$ with $H_0$ obtained from SH0ES data and the results are presented in figure (\ref{HTdiffalpha}).
Upon analysing the results, we find that $H_T$ is not influenced by the quantum gravity effect. Therefore we can conclude that quantum gravity does not play any role in phase transition. This is in agreement with the equation (\ref{bj2}). \\

We observe that quantum gravity can have an effect on $H_I$ whose signature can be traced in the reheating stage. Hence we investigate the quantum gravity effect on the duration of the reheating and reheating temperature with Planck 2018 data to substantiate the impact of quantum gravity on $H_I$.


\subsection{Quantum gravity effect on reheating}

We have seen that the number of e-folding of reheating and the reheating temperature are depended on the scalar spectral index and the amplitude of the scalar power spectrum (see equation (\ref{N_re}) and (\ref{T_re})). But, these quantities get modified in the presence of quantum gravity (see equation (\ref{nsqg}) and (\ref{asqg})) and have an effect on $N_{re}$ and $T_{re}$. Therefore using equation (\ref{nsqg}) and (\ref{asqg}) in equation (\ref{N_re}) and (\ref{T_re}), the quantum gravity corrected reheating e-folding number is obtained as
\begin{equation}
\begin{split}
N_{re} = N_{re, cl} + N_{re, qg}  = & \Bigg\{ \frac{4}{1-3\text{w}}\bigg[ 61.6 - \frac{1}{4} \ln \bigg( \frac{6912 \  \Omega_{\Lambda}^2  \ (1-n_{s,cl})^2}{32768 \ \pi^2 \ A_{s,cl}} \bigg)  - \frac{2}{1-n_{s,cl}} \bigg] \Bigg\}  \\ & + \Bigg\{ \frac{4}{1-3\text{w}}  \bigg[ \frac{2}{1-n_{s,cl}} - \frac{1}{4} \ln \bigg(    \frac{1- \frac{n_{s,qg}}{1-n_{s,cl}}}{1+\frac{A_{s,qg}}{A_{s,cl}}}     \bigg) - \frac{2}{1-n_{s,cl}-n_{s,qg}} \bigg] \Bigg\}.
\end{split}
 \end{equation}
 \begin{figure}[t]
   \centering
   \includegraphics[width=1\linewidth]{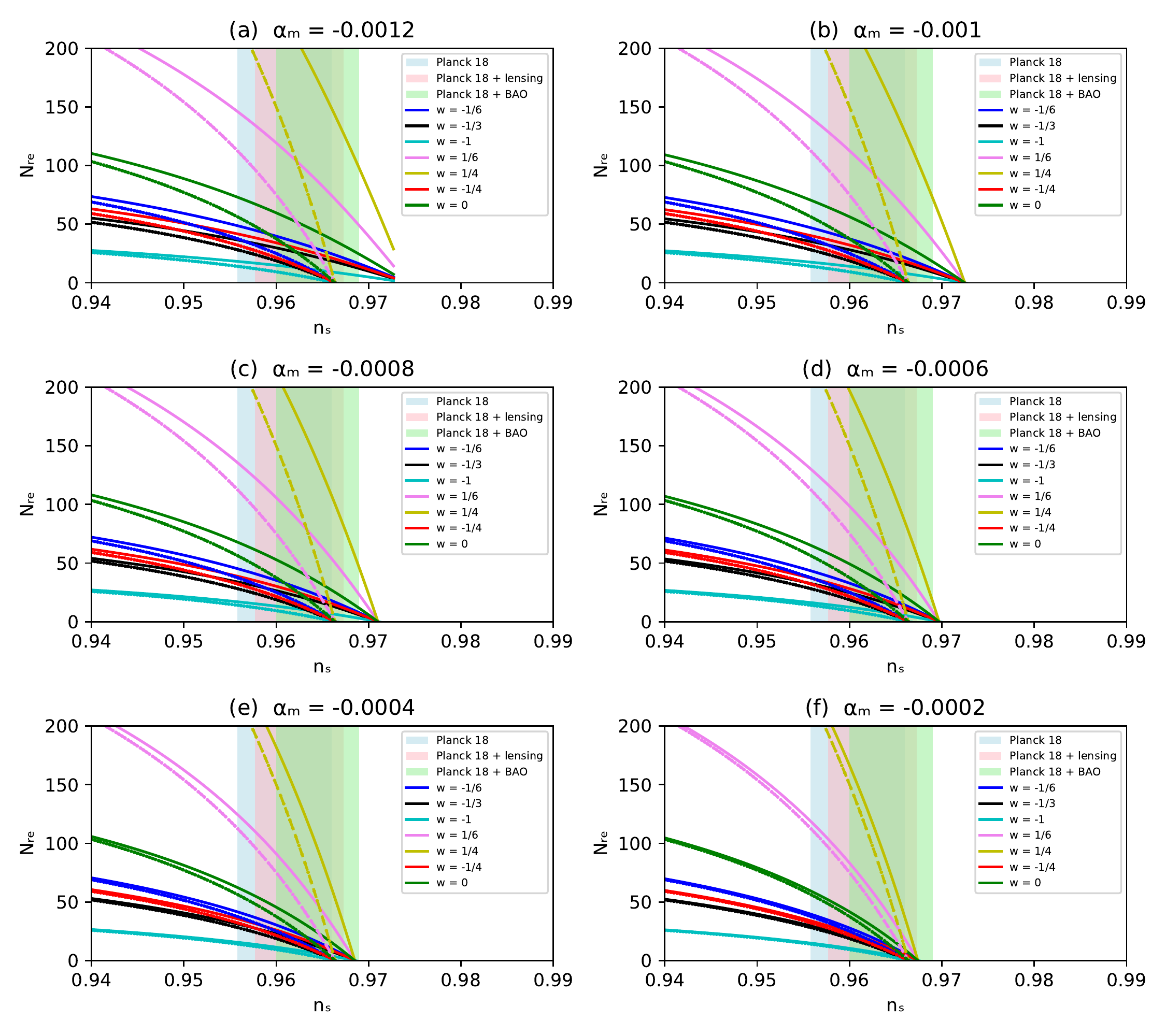}
      \caption{Quantum gravity effect on the reheating e-folding number ($N_{re}$) for a range of scalar spectral index ($n_s$) for different values of equation of state (w) and higher dimensional operator $\alpha_m$ with Planck 2018 data.}
      \label{rehn}
\end{figure}
And similarly, the quantum gravity corrected reheating temperature can be written as 
\begin{equation}
\begin{split}
T_{re} = T_{re, cl} \times T_{re, qg} = & \Bigg\{  \bigg(\frac{43}{11}\bigg)^{\frac{1+\text{w}}{3\text{w}-1}} \ \bigg( \frac{135}{2} \bigg)^{\frac{-1}{3\text{w}-1}}  \ {g_*}^{\frac{-\text{w}}{3\text{w}-1}} \ \bigg( \frac{\sqrt{2} \ \pi \  a_0 \ T_0 }{k} \bigg)^\frac{3(1+\text{w})}{3\text{w}-1} \\ 
&\times m_{pl}^{\frac{3\text{w}+1}{3\text{w}-1}}  A_{s,cl}^{\frac{1+3\text{w}}{2(3\text{w}-1)}} \ (1-n_{s,cl})^{\frac{1}{2}} \  e^{\frac{-6(1+\text{w})}{(1-n_{s,cl})(3\text{w}-1)}}  \Bigg\} \times \Bigg\{  \bigg( 1+ \frac{A_{s,qg}}{A_{s,cl}} \bigg)^{\frac{1+3\text{w}}{2(3\text{w}-1)}} \\ &\times  \sqrt{ 1+ \frac{n_{s,qg}}{(1-n_{s,cl})}}  e^{- \bigg[ \frac{1}{1- \big( \frac{n_{s, qg}}{1-n_{s, cl}} \big)} \bigg]} \Bigg\} .
\end{split}
\end{equation}
\begin{figure}[t]
   \centering
   \includegraphics[width=1\linewidth]{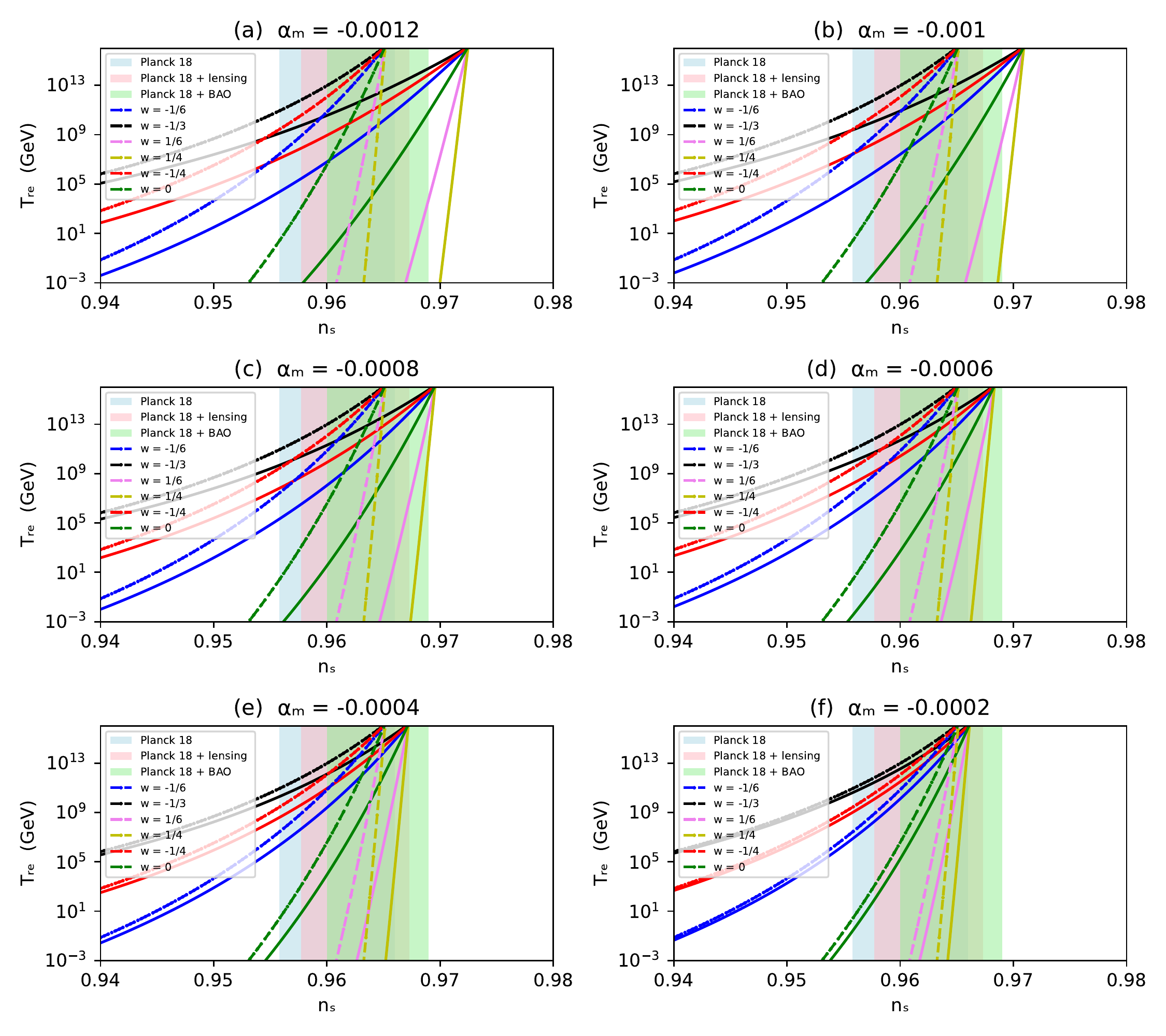}
      \caption{Role of quantum gravity on the reheating temperature ($T_{re}$) for a range of scalar spectral index ($n_s$) for different values of equation of state (w) and higher dimensional operator ($\alpha_m$) with Planck 2018 data.} 
      \label{reht}
\end{figure}
We study the reheating e-folding number for different values of equation of state and $\alpha_m$ with a range of scalar spectral index obtained from Planck 2018 data. The results are presented in figure (\ref{rehn}). For $\alpha_m$ $\leq$ $-$ 0.0012 the reheating e-folding numbers representing the various equation of state do not converge to a single $n_s$ value indicating that the quantum gravity effect is equation of state dependent. But this is unphysical and unlikely to happen as the particles are not yet produced before reheating and therefore the corresponding $\alpha_m$ values are ruled out (see figure \ref{rehn} (a)). Moreover, the obtained $N_{re}$ values are blue tilted with respect to the Planck 2018 bounds. For some values of $\alpha_m$, the $N_{re}$ values converge at higher values of $n_s$ showing that the different w agree with each other, but the corresponding $\alpha_m$ values are ruled out as they converge outside the Planck 2018 bounds (see figure \ref{rehn} (b), (c) and (d)). For higher values of $\alpha_m$ the $N_{re}$ values converge within the Planck 2018 bounds and their corresponding $\alpha_m$ values are favourable (see figure \ref{rehn} (e) and (f)). But in all these cases $N_{re}$ after quantum gravity correction is found to be higher than $N_{re,cl}$ indicating that the quantum gravity effect can elongate the reheating period. The prolonged reheating suggests that it has taken much more time to attain the reheating temperature or in other words it must have started with a lower temperature of reheating. To check this, we examine the variation of reheating temperature for a range of values of scalar spectral index and for different values of equation of state and HDO. The results are presented in figure (\ref{reht}). It is observed that due to the quantum gravity effect, the reheating begins with a lower temperature which implies that it takes more time to achieve the reheating temperature (reheating is prolonged) in consistent with the previous results obtained for $N_{re}$. We can see that for some $\alpha_m$ values $T_{re}$ converges to a single value at a particular $n_s$ which is well within the Planck 2018 bounds suggesting that $T_{re}$ is independent of the equation of state (see figure \ref{reht} (d), (e) and (f)). From the study and analysis of the present work, it is evident that quantum gravity can decrease the temperature during the onset of reheating thereby increasing the duration of reheating. Therefore the role of quantum gravity cannot be ignored while addressing the Hubble tension.


\section{Discussions and conclusions} 

The discrepancy existing in the present value of the Hubble parameter obtained from Planck 2018 and SH0ES data creates a tension in understanding the universe and therefore resolving it is vital in the field of cosmology. In spite of many attempts to resolve the Hubble tension, the lack of a satisfactory solution motivates us to address the Hubble tension starting from the inflationary stage of the universe with novel effects such as the quantum gravity effect. In the standard framework, the single scalar field responsible for inflation is supposed to be insensitive to quantum gravity effect. However combining CMB results with Lyth bound suggests that the inflationary field can have energy higher than the Planckian energy scale. Further, CMB results favour multi fields over single field in the inflationary scenario. Among the multi fields inflationary models, the hybrid model with two scalar fields received much attention, where one field drives the inflation and the other is responsible for phase transition. This gives rise to the necessity of considering the Hubble parameter during inflation and phase transition separately. Therefore we incorporate quantum gravity in the hybrid inflationary model using the effective field theory approach to investigate its effect on $H_0$ through $H_I$ and $H_T$. We observe that quantum gravity can have an impact on $H_I$ whereas it does not reflect on $H_T$. We find that the effect of quantum gravity during inflation can increase $H_0$ thus accounting for the different observed $H_0$ values thereby suggesting that quantum gravity may be a viable solution to the Hubble tension.  
Since quantum gravity is playing a prominent role during inflation the chances of its manifestation on the subsequent stages of reheating cannot be ruled out. The footprints of quantum gravity on $H_I$ can be substantiated by studying its aftereffect on reheating through reheating e-folding number and reheating temperature with the scalar spectral index coming from Planck 2018 data. Finally, we show that the quantum gravity effect during inflation can influence the reheating by lowering the initial temperature of reheating thereby delaying the reheating process. We may conclude that the quantum gravity effect during inflation is inevitable in addressing Hubble tension. Further, the results of the present study may be useful in validating inflationary model. \\

The present work is mainly to resolve the Hubble tension with quantum gravity effect on the hybrid inflationary model. Results and observations of this study can be re-examined with other inflationary models, which can help not only in resolving the Hubble tension but also in validating inflationary model because resolving the Hubble tension relies on the underlying inflationary model. Therefore the Hubble tension cannot be studied in isolation but it has to be in conjunction with the validation of inflationary model. At a glance validating inflationary model may not have any direct connection to Hubble tension because it relies chiefly on the tensor to scalar ratio. But in the light of Hubble tension, an appropriate inflationary model has to be considered to account for the observed values of $H_0$. Therefore validation of inflationary model can result in resolving the Hubble tension and vice versa. In other words, validation of inflationary model and resolution of Hubble tension can complement each other. The results of the present work are useful in resolving the Hubble tension as well as validating inflationary model along with quantum gravity.

\section*{Acknowledgement}

AB acknowledges the financial support of Prime Minister's Research Fellowship (PMRF ID : 3702550).



\begin{thebibliography}{99}

\bibitem{henri}  Henrietta S Leavitt, \emph{1777 Variables in the Magellanic Clouds},  \emph{Harvard Obs.Annals.} {\bf 15} (1908) 87.

\bibitem{vesto}  Vesto M Slipher, \emph{Nebulae}, \emph{Proc. Am. Philos. Soc.} {\bf56} (1917) 403.

\bibitem{lemaitre} G Lema\^{i}tre, \emph{A Homogeneous Universe of Constant Mass and Increasing Radius Accounting for the Radial Velocity of Extra-galactic Nebulae}, \emph{Mon. Notices Royal Astron. Soc.} {\bf 91} (1931) 483.

\bibitem{iau} \emph{Resolutions presented to the $XXX^{th}$ General Assembly of the International Astronomical Union}, \emph{https://www.iau.org/news/pressreleases/detail/iau1812} (2008). 

\bibitem{hubble} Edwin Hubble, \emph{A relation between distance and radial velocity among extra-galactic nebulae}, \emph{Proc. Natl. Acad. Sci.} {\bf 15} (1929) 168. 

\bibitem{devalentino} Eleonora Di Valentino et al., \emph{In the realm of the Hubble tension—a review of solutions}, \emph{Class. Quantum Grav.} {\bf 38} (2021) 153001.

\bibitem{aghanim} Aghanim et al., \emph{Planck 2018 results. VI. Cosmological parameters}, \emph{Astron. Astrophys.} {\bf A6} (2020) 641.

\bibitem{shoes} Riess Adam G et al., \emph{A Comprehensive Measurement of the Local Value of the Hubble Constant with 1 $Km\  s^{-1}\  MPc^{-1}$ Uncertainty from the Hubble Space Telescope and the SH0ES Team}, \emph{Astrophys. J.} {\bf 934} (2022) L7.

\bibitem{ldl} Edvard Mortsell et al., \emph{The Hubble Tension Revisited: Additional Local Distance Ladder Uncertainties}, arXiv:2106.09400v3.

\bibitem{ede} Vivian Poulin et al., \emph{Cosmological implications of ultralight axion like fields}, \emph{ Phys. Rev. D.} {\bf 98} (2018) 083525.

\bibitem{jordan} Tiziano Schiavone and Giovanni Montani, \emph{f(R) gravity in the Jordan Frame as a Paradigm for the Hubble Tension}, arXiv:2211.16737v1.

\bibitem{serg} S Nojiri, S.D Odintsov and V K Oikonomou, \emph{Integral F(R) gravity and saddle point condition as a remedy for the $H_0$-tension}, \emph{Nucl. Phys. B} {\bf 980} (2022) 115850.

 \bibitem{deg} Alvaro S. de Jesus et al., \emph{The Hubble Rate Trouble: An Effective Field Theory of Dark Matter}, arXiv:2212.13272v1.

\bibitem{vacint} Li-Yang Gao et al., \emph{Dark energy and matter interacting scenario can relieve $H_0$ and $S_8$ tensions}, arXiv:2212.13146v1.

\bibitem{ssmb} Celia Escamilla-Rivera and Ruben Torres Castillejos, \emph{$H_0$ Tension on the Light of Supermassive Black Hole Shadows Data}, arXiv:2301.00490v1.

\bibitem{gs} Ish Gupta, \emph{Using Gray Sirens to Resolve the Hubble-Lemaître Tension}, arXiv:2212.00163v1.

\bibitem{ellipse} Paolo Cea, \emph{The Ellipsoidal Universe and the Hubble tension}, arXiv:2201.04548.

\bibitem{35} Zachary G. Lane, Antonia Seifert, Ryan Ridden-Harper, Jenny Wagner, and David L. Wiltshire, \emph{Cosmological foundations revisited with Pantheon+}, arXiv:2311.01438v1. 

\bibitem{nf} P. K. Aluri, et al., \emph{Is the observable Universe consistent with the cosmological principle?}, \emph{Class. Quant. Grav.}, {\bf40} (2023) 094001.

\bibitem{ct} D. L. Wiltshire, \emph{Cosmic structure, averaging and dark energy}, arXiv:1311.3787.

\bibitem{ts}D. L. Wiltshire, \emph{Cosmic clocks, cosmic variance and cosmic averages}, \emph{New J. Phys.}, {\bf9} (2007) 377.

\bibitem{lyth} D H Lyth, \emph{What would we learn by detecting a gravitational wave signal in the cosmic microwave background anisotropy?} \emph{Phys. Rev. Lett.} {\bf 78} (1997) 1861.

\bibitem{linde} Andrei Linde, \emph{ Hybrid Inflation}, \emph{Phys. Rev. D.} {\bf 49} (1994) 748.

\bibitem{bourakadi} K. El Bourakadi,\emph{ Hubble tension and Reheating: Hybrid Inflation Implications}, arXiv:2208.01162v1.

\bibitem{guth} Alan H. Guth, \emph{ Inflationary universe : A possible solution to the horizon and flatness problems}, \emph{Phys. Rev. D.} {\bf 23} (1981) 347.

\bibitem{encyclo} Jerome Martin, Christophe Ringeval, and Vincent Vennin, \emph{ Encyclopædia inflationaris}, \emph{Phys. Dark Universe.} {\bf 5} (2014) 75.

\bibitem{waterfall} Hideo Kodama et al., \emph{ On the Waterfall Behaviour in Hybrid Inflation}, \emph{Prog. Theor. Phys.} {\bf 126} (2011) 331.

\bibitem{bdv} T. S. Bunch and P. C. W. Davies, \emph{Quantum Field Theory in De Sitter Space : Renormalization by Point-Splitting}, \emph{Proc. R. Soc. A.} {\bf 360} (1978) 117.

\bibitem{ups} Vivian Poulin, Tristan L. Smith and Tanvi Karwal, \emph{The Ups and Downs of Early Dark Energy solutions to the Hubble tension: a review of models, hints and constraints circa 2023}, arXiv:2302.09032v2.

\bibitem{kami} Marc Kamionkowski1 and Adam G. Riess, \emph{The Hubble Tension and Early Dark Energy}, arXiv:2211.04492.

\bibitem{akhilesh}  Akhilesh Nautiyal, \emph{ Reheating constraints on tachyon inflation}, \emph{Phys. Rev. D.} {\bf 98} (2018) 103531.

\bibitem{kaz1} M Kawasaki, K Kohri and Naoshi Sugiyama, \emph{Cosmological Constraints on Late-time Entropy Production}, \emph{Phys.Rev.Lett.} {\bf 42} (1999) 4168.

\bibitem{kaz2} M Kawasaki, K Kohri and Naoshi Sugiyama, \emph{MeV-scale Reheating Temperature and Thermalization of Neutrino Background}, \emph{Phys. Rev. D.} {\bf 62} (2000) 023506.

\bibitem{historical} A.D Linde, \emph{ A New Inflationary universe scenario: a possible solution of the horizon, flatness, homogeneity, isotropy and primordial monopole problems}, \emph{Phys. Lett.} {\bf 108B} (1982) 389.

\bibitem{tachyonic} Juan Garcia-Bellido, \emph{ Tachyonic preheating and spontaneous symmetry breaking}, arXiv:hep-ph/0106164.

\bibitem{parametric} Juan Garcia-Bellido et al., \emph{ Preheating in Hybrid Inflation}, \emph{Phys. Rev. D.} {\bf 57} (1998) 6075.

\bibitem{parametric2} Kofman et al., \emph{ Towards the theory of reheating after inflation}, \emph{Phys. Rev. D.} {\bf 56} (1997) 3258.

\bibitem{instant} Gary Felder et al., \emph{ Instant preheating }, \emph{Phys. Rev. D.} {\bf 49} (1999) 123523.
 
\bibitem{gauss} K. El Bourakadi et al., \emph{ Gravitational waves from preheating in Gauss–Bonnet inflation}, \emph{Eur. Phys. J. C.} {\bf 81} (2021) 1.

\bibitem{kosar} Asadi Kosa and Kourosh Nozari, \emph{ Reheating constraints on a two-field inflationary model}, \emph{ Nucl. Phys. B.} {\bf 949} (2019) 114827.

\bibitem{n_s} K. El Bourakadi, \emph{ Preheating and Reheating after Standard Inflation}, arXiv:2104.10552v2.

\bibitem{donos} J.F Donoghue, \emph{ The effective field theory treatment of quantum gravity}, \emph{AIP Conf. Proc.} {\bf1483} (2012) 73.

\bibitem{calmet} X. Calmet, \emph{ Effective theory for quantum gravity}, \emph{Int. J. Mod. Phys. D.} {\bf 22} (2013) 1342014. 

\bibitem{excursion} X. Calmet and V. Sanz, \emph{ Excursion into quantum gravity via inflation}, \emph{Phys. Lett. B.} {\bf 737} (2014) 12.

\bibitem{sur} P K Suresh, \emph{ Signature of the Quantum Gravity on the CMB },  \emph{Mod. Pers. Theor. Phys} Springer Nature (2021) 23.

\end{thebibliography}
\end{document}